\documentclass[epj,referee]{svjour}

\usepackage{graphics}
\begin{document}
\title{ Probing helium interfaces with light scattering : from fluid mechanics
to statistical physics}
\author{{P. E.}{Wolf}\inst{1},  {F.}{Bonnet}\inst{1},  {S.}{Perraud}
\inst{1,2}, {L.}{Puech}\inst{1}, {B.}{Rousset} \inst{2}, {P.}{Thibault}
\inst{2}
}                     
\authorrunning {P.E. Wolf \it {et al} \rm}%
\titlerunning {Probing helium interfaces with light scattering}%
\offprints{pierre-etienne.wolf@grenoble.cnrs.fr}          
\institute{Institut N\'eel, CNRS-UJF, BP 166,
38042 Grenoble-Cedex 9, France\and CEA
Grenoble-UMRE 9006, INAC, Service des Basses Temp\'eratures, 17 
rue des Martyrs, 38054 Grenoble-Cedex 9, France}
\date{Received: June 17 th / Revised version: date}
%
\abstract{ We have investigated the formation of helium droplets in
two physical situations.  In the first one, droplets are atomised from
superfluid or normal liquid by a fast helium vapour flow.  In the
second, droplets of normal liquid are formed inside porous
glasses during the process of helium condensation.  The
context, aims, and results of these experiments are reviewed, with
focus on the specificity of light scattering by helium.  In
particular, we discuss how, for different reasons, the closeness to
unity of the index of refraction of helium allows in both cases to
minimise the problem of multiple scattering and obtain results which 
it would not be possible to get using other fluids.
\PACS{
      {67.25.bf}{Normal phase of 4He : Transport, hydrodynamics}   \and
      {67.25.bh}{Films and restricted geometries}   \and
     {47.80.Jk}{Flow visualization and imaging}    \and
     {47.55.-t}{Multiphase and stratified flows}    \and
     {67.25.dg}{Superfluid phase of 4He : Transport, hydrodynamics, and superflow} 
     } 
} 
\maketitle
\section{Introduction}
\label{intro}
Light scattering is a widely used tool to investigate problems in soft
matter physics.  In many instances, multiple scattering turns out to
be a problem.  During the last twenty years, significant progresses in
dealing with and even exploiting multiple scattering have been made.
Locating objects inside turbid media has been shown to be possible by
measuring the static\cite{denOuter:93} or dynamic scattered intensity.
In particular, the so-called dynamic wave
spectroscopy\cite{Maret1987,Pine1988} enables the visualisation of
flows\cite{Heckmeier1996} or the location of absorbing objects
\cite{Heckmeier1997} inside multiple-scattering liquids.  However,
imaging under multiple scattering conditions is challenging and
information provided by such techniques remains limited compared to
those obtained by single scattering.  This explains why a great deal
of effort has been devoted to the suppression or the minimisation of
multiple scattering in dense soft-matter systems.  A first example is
the study of the glass transition in colloidal suspensions of silica
and copolymer spherical particles\cite{VanMegen1998}.  Here, the
solvent index of refraction is matched to that of the copolymer
particles, so that scattering only arises from the silica particles.
Their volume fraction is small enough (2\%) to study their dynamics
using quasi-elastic light single scattering (the small amount of
multiple scattering being eliminated by two-colour cross-correlation
\cite{Segre1995}).  A second example is the study of phase separation
of binary mixtures in the presence of a disordered environment,
provided by confining the mixture inside a porous
medium\cite{Maher1984,Dierker1991,Frisken1995}.  Near the critical
point, the density fluctuations give rise to multiple scattering which
is minimised by careful choice of the refractive index of the fluids
with respect to that of the substrate (e.g. isobutyric acid and water
in silica aerogels \cite{Frisken1995}).

\begin{figure}
\center
\resizebox{1\columnwidth}{!}{
  \includegraphics{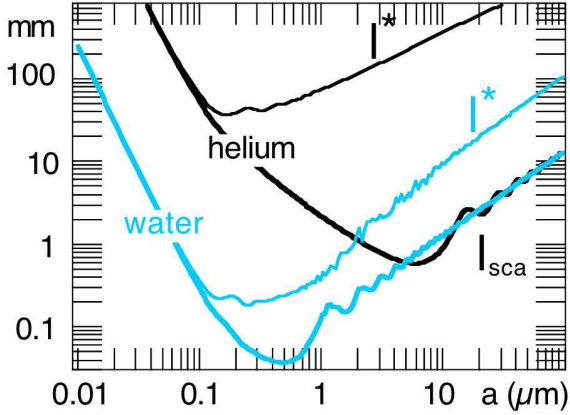}
} \caption{Scattering and transport mean free paths of light $l_{sca}$
and $l^{*}$ (in mm), for uncorrelated helium and water spherical
droplets of radius $a$ (index contrast 1.025 and 1.33), at a volume
fraction of 5.10$^{-3}$.  The wavelength is 632 nm.  }
\label{figurelpm}       
\end{figure}

Such an index matching does not occur when one considers a diphasic
pure fluid.  In this case, the optical contrast between the vapour and
the liquid phases is solely fixed by the ratio of their densities and
cannot be tuned.  This makes the optical study of diphasic systems
prone to the occurrence of multiple scattering.  From this point of
view, helium is a remarkable exception.  Due to its small size and
closed shell electronic structure, its polarisability is very low and
its index of refraction close to 1.  For liquid helium below 2 K
(density of 145 kg/m$^3$), the dielectric permitivity is
1.057\cite{McCarty}, and the index of refraction 1.028.  This implies
that the scattering cross-section for scatterers much smaller than the
light wavelength is very small, so that multiple scattering barely
occurs.  The situation for larger scatterers is more subtle.  In the
regime of geometrical optics, the scattering cross section is
approximately twice the geometrical cross-section (diffraction and
reflection/refraction contributing nearly equally\cite{vandehulst}),
so that, at a given volume fraction, the scattering mean free path decreases with
the particle diameter (figure~\ref{figurelpm}), and reaches a minimum
for a particle radius of about 10 $\mu$m.  Even for moderate volume
fractions (10$^{-3}$), it can become smaller than a light path length
of 1~cm, giving rise to multiple scattering.  However, the small
refractive index of helium implies that the scattering is nearly
forward in this case.  Figure~\ref{figurelpm} shows that, for droplets
of liquid helium, the transport mean free path $l^{*}$ is then more
than two orders of magnitude larger than the scattering mean free
path $l_{sca}$.  As we will show, this peculiar situation makes imaging possible
even through samples much larger than $l_{sca}$.

In this paper, we will describe two physical situations where we take
full advantage of the low refractive index of helium.  The first one
is the study of atomisation of normal or superfluid liquid helium by a
fast vapour flow.  The droplets created through this process range
from microns to tens of microns, so that geometrical optics applies.
Even when the droplets density becomes large, observation at
appropriate angles allows to determine their spatial distribution, and
to analyse the influence of the physical characteristics of helium on
the atomisation process.  The second situation is the condensation and
evaporation of helium inside porous silica glasses of different
topologies, Vycor and silica aerogels.  Both processes involve
fluctuations of the helium density on a submicronic scale, which can
be probed by spatially resolved light single scattering.  This
provides information on the size of the microscopic heterogeneities
and their macroscopic distribution, which can both be compared to the
predictions of the existing models for the condensation and
evaporation processes.

\section{The case of large scatterers : Atomisation of liquid helium  }
\label{Atomisation}
\subsection{The Cryoloop experiment}\label{atom:cryoloop}
Blowing a fast enough gas stream parallel to the free surface of a
liquid results in the liquid atomisation into a spray of droplets.
The Cryoloop experiment aims at studying this process for diphasic
helium in a horizontal pipe.  The liquid can be either normal or
superfluid, depending on the temperature, and the gas is helium vapour
at the saturated vapour pressure.  The interest of this experiment is
two-fold.  First, diphasic flows of helium can be used for
refrigeration purposes, an example (below the atomisation threshold)
being the cooling of the superconducting magnets of the Large Hadron
Collider, and it is important to assay the conditions and consequences
of atomisation.  Second, the physical parameters relevant for
atomisation are very different for diphasic helium and for the widely
studied air-water system (see table~\ref{tableHe}), which offers an
opportunity to experimentally test the influence of these parameters
on the atomisation process.

\begin{table}
\caption{Comparison of the liquid and vapour densities 
($\rho_{L}$, $\rho_{V}$) and 
viscosities ($\eta_{L}$, $\eta_{V}$), and of the liquid-vapour interfacial energy ($\sigma$) for the 
diphasic helium and water-air systems; data for helium from 
refs.~\cite{VanSciver98} and  (for $\sigma$) \cite{Iino-JLTP85}.}
\label{tableHe}       
\begin{tabular}{llllll}
\hline\noalign{\smallskip}
fluids & $\rho_{L}$  &$\rho_{V}$ &$\eta_{L}$ 
 &$\eta_{V}$ &$\sigma$  \\
 &  (kg/m$^3$) &(kg/m$^3$) & 
($\mu$Pa.s) &($\mu$Pa.s) & (mJ/m$^2$) \\
\noalign{\smallskip}\hline\noalign{\smallskip}
He 1.8 K & 145 & 0.48& 1.3 & 0.43& 0.316 \\
He 2.0 K & 145 & 0.79& 1.5 & 0.52 & 0.301 \\
He 2.2 K & 146 & 1.25& 2.6 & 0.57& 0.284 \\
He 2.6 K & 144 & 2.55& 3.3 & 0.69& 0.253 \\
Water/air & 1000 & 1.2& 1000 & 17& 70 \\
\noalign{\smallskip}\hline
\end{tabular}
\end{table}

The Cryoloop facility is installed at the Service des Basses
Temp\'eratures at CEA-Grenoble.  A specially developed
refrigerator\cite{Roussel05} with large cooling power (400~W at 1.8~K)
is used to deliver a flow of liquid of up to approximately 20~g/s (i.e. more than
400~litres of liquid per hour) to one end of a nearly horizontal, 10~m
long and 40~mm inner diameter, pipe.  This liquid is partly evaporated
by a heater in order to obtain downstream an essentially stratified
diphasic flow in which the bulk liquid moves along the bottom of the
pipe, and the vapour moves above the liquid.  At the end of the pipe,
the remaining liquid is evaporated, and the resulting vapour is pumped
through cold compressors followed by a room temperature pump, which
sets the pressure, hence the temperature, inside the pipe.

At a given applied power (i.e. a given mass flow rate of vapour), the
vapour velocity depends on the cross-section of the pipe accessible to
the vapour and on its density.  In our experiments, the former is
essentially the whole cross-section of the pipe, so that the ultimate
maximal (mean) vapour velocity decreases with increasing temperature (since
the saturated vapour pressure, hence the vapour density, increases with
temperature).  In order to obtain atomisation, a part of the total
flow has to be kept liquid, which makes the practical maximal velocity
to depend on the amount of liquid left.  Typically, for a liquid
height of 3.5~mm (corresponding to a wetted fraction of the pipe of
20\%), the maximal velocity decreases from 18~m/s at 1.8~K to 6~m/s at
2.5~K.

\subsection{Visualisation of atomisation}\label{atom:visualisation}
\begin{figure}
\center
\resizebox{0.9\columnwidth}{!}{
  \includegraphics{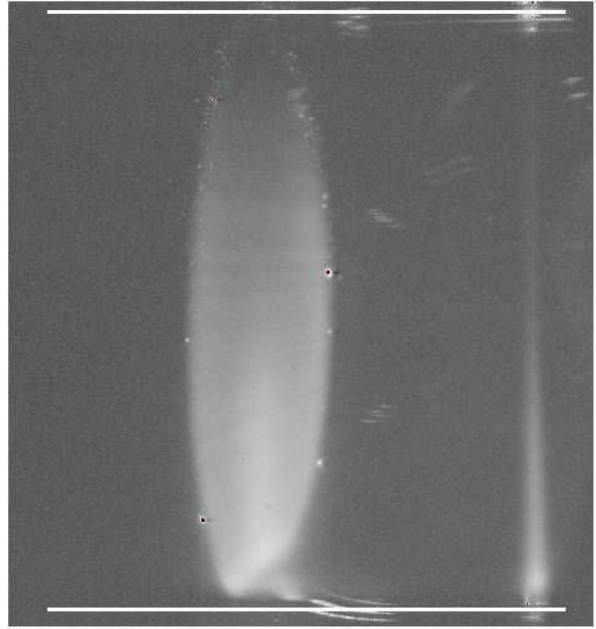}
}
\caption{Helium mist generated by atomisation in a 
4~cm diameter pipe (log-gray scale). The inner walls of the pipe correspond to the white 
bars. The mist is visualised using illumination by a 
laser sheet
at 15$^{\circ}$ from the CCD direction of observation (left), and by a 
vertical laser (right). The liquid-vapour interface is visible as a 
brighter region in the bottom of the pipe.  
}
\label{fogpicture}       
\end{figure}

In order to observe the droplets created by the atomisation process,
the stainless steel pipe is replaced along a length of 13~cm by a
Pyrex tube of identical inner diameter.  This optical section is
located close to the end of the pipe to ensure that the observed spray
is fully developed.  Viewports through the cryostat containing the
pipe allow to illuminate and to detect the spray using
room-temperature optical components\cite{DiMuoio01}.  The spray distribution across
the whole cross-section of the pipe can be observed by illuminating
the pipe with a laser sheet propagating horizontally perpendicular to
the pipe.  A CCD located opposite to the laser source with respect to
the pipe is used to image the illuminated section under a 15$^{\circ}$
angle, as shown in figure~\ref{fogpicture}.  With the same CCD, we
also observe the light scattered at 90$^{\circ}$ from a laser beam
propagating vertically along the pipe diameter, which allows to
compare scattering at two angles along this particular light path.

\begin{figure}
\center
\resizebox{\columnwidth}{!}{%
  \includegraphics{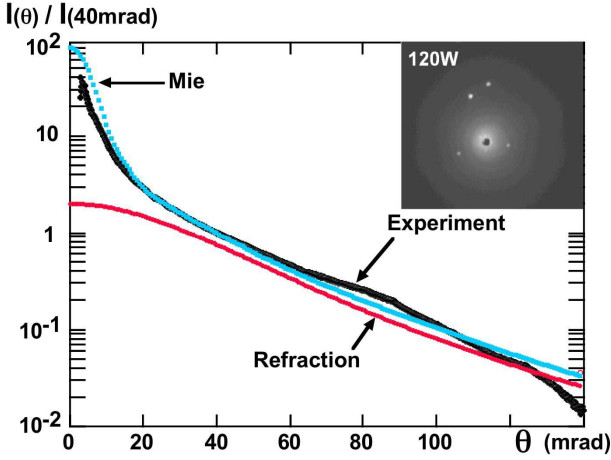}
} \caption{Angular dependence of the intensity scattered by the mist,
as measured on a screen perpendicular to the incident laser beam.  The
unscattered beam passes through the hole at the center of the picture.
Measuring its intensity gives the scattering mean free path, averaged
over the horizontal diameter of the pipe.  White spots correspond to
reflections on the windows.  The scattered intensity does not depend
on the azimuthal angle $\phi$, which means that the droplets are not
elongated by the gas flow.  Averaging this intensity over
$\phi$ gives the thick curve.  Above 40 mrad, it follows
the prediction of geometrical optics, showing that the droplet size is
larger than one micron.  The rise at smaller angles is due to
diffraction and compares well to the Mie prediction for an exponential
distribution of diameters with characteristic diameter 20 $\mu$m.}
\label{angulardep}       
\end{figure}

A second optical access opens to a stainless steel portion of the pipe
through small flat windows, centred on the pipe horizontal
median plane.  Shining a horizontal laser beam through these windows
and visualising the scattered intensity on a screen located on the
other side of the pipe allows to determine the angular dependence of
the scattered intensity in a cone of about 10$^{\circ}$ around the
forward direction.  This enables to test the scattering regime, i.e.
the size of the droplets, on the median plane of the pipe.
Figure~\ref{angulardep} shows that the scattering is strongly peaked
in the forward direction.  The measured angular dependence is that
predicted by geometrical optics for scattering angles larger than
40~mrad, with a steep rise at smaller angles, associated with
diffraction.  The observation of refraction shows that the average
diameter of droplets is larger than one micron.  This is confirmed by
both the angular range over which diffraction occurs and measurements
with a Phase Doppler Particle Analyser (PDPA), which show that the average
diameter is a few tens of microns, for all hydrodynamic conditions. 

Being in the regime of geometrical optics implies that the scattered
intensity (outside the diffraction peak) is directly proportional to
the interfacial area $\Sigma$, per unit total volume, of the droplets.
This also holds for the scattering mean free path, since the total
scattering cross section is twice the geometrical one (more precisely,
there are oscillations with size, due to interference effects, of both
quantities, but they are washed out by the average over the droplets
size distribution).  By comparing, for given conditions, the image of
the mist with the attenuation of the horizontal beam used for
measuring the angular dependence, we can determine the coefficient
between the brightness of the images and $\Sigma$, and thus, extract
from any image the spatial dependence of $\Sigma$ (or, equivalently,
of the local mean free path of light $l_{sca}$=2/$\Sigma$).

\begin{figure}
\center
\resizebox{0.9\columnwidth}{!}{%
  \includegraphics{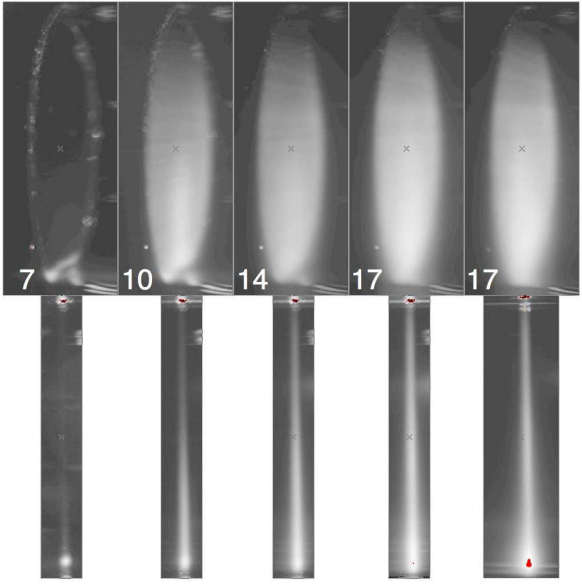}
}
\caption{Mist for velocities (averaged over the pipe section) of 7, 
10.5, 14, 17.2 and 17.6 m/s, at a temperature of 1.8~K and a liquid 
level corresponding to the bottom 20\% of the pipe perimeter. 
The images are averages of several tens of pictures with 30~ms 
exposure time (10~ms for the two pictures above 17~m/s). A log-gray 
scale is used both for sheet and vertical laser illumination. }
\label{picturesV}       
\end{figure}

\begin{figure}
\center
\resizebox{1\columnwidth}{!}{%
   \includegraphics{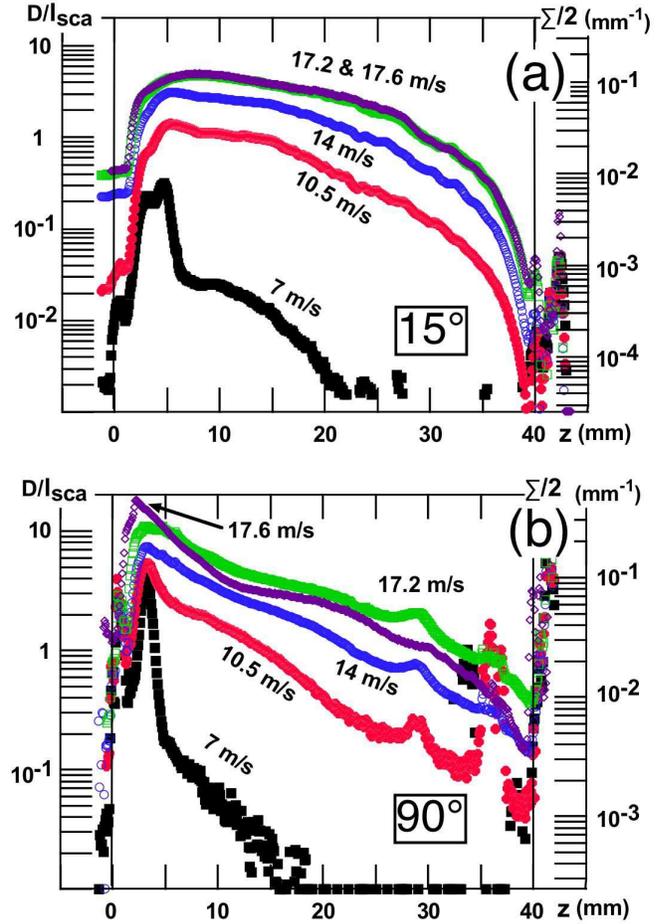}
   }
\caption{Mist stratification~:~Profiles of interfacial density deduced from
figure~\ref{picturesV} for sheet (15$^{\circ}$, top) and vertical laser 
(90$^{\circ}$, bottom) 
illumination.  The left scales gives the corresponding ratio
of the pipe diameter $D$ to the local mean free path $l_{sca}$. The 
right scale is $1/l_{sca}=\Sigma /2$.  The increase of
signal beyond 40~mm is due to scattering from the pipe walls.  The
peaks above 20~mm for laser illumination arise from similar scattering
of the beam reflected by the bottom of the pipe. }
\label{profilesV}       
\end{figure}

Figure~\ref{picturesV} displays pictures separately recorded with
laser sheet and vertical laser illumination for increasing vapour
velocities, at constant bulk liquid level and vapour temperature.  The
obvious strong increase of atomisation with vapour velocity is
quantified in figure~\ref{profilesV}, which shows the dependence of
the interfacial area on the height $z$ inside the pipe, as deduced
from both modes of illumination.  The curves show that the mist is
stratified over a characteristic height $H_{char}$ which increases with
the vapour velocity.  This stratification results from the competition
between the diffusion of the droplets due to the vapour turbulence and
their gravitational fall through the viscous vapour\cite{Paras91b}.
The decrease of stratification with increasing vapour velocity can then
be ascribed to the increase of the droplets turbulent diffusivity or
to a decrease of their settling velocity due to a smaller size.
Beyond this common behaviour, a striking difference between the two
modes of illumination is that, in the bottom part of the tube ($z<15$
mm), the interfacial intensity deduced from sheet illumination
saturates while that deduced from the vertical laser still grows.
This discrepancy occurs for values of the mean free path
smaller than the pipe diameter (which is close to the distance
travelled by the laser sheet, except very near the bottom of the pipe).
This suggests that multiple scattering is involved.  
The question then arises of how the path of the vertical laser can be
imaged in a region where the mean free path is as small as a tenth of
the pipe diameter.  A qualitative answer is that, because the
transport mean free path for helium droplets is so much larger than
the scattering mean free path, the light scattered at a large angle
involves only one scattering event at (approximately) this angle,
between two series of scattering events at small angle.  However, it
is not obvious whether this mechanism is more likely than a series of
small-angle scattering events.  In fact, as we now discuss, this
depends not only on the ratio of the transport mean free path $l^{*}$
to the scattering mean free path $l_{sca}$, but also \cite{Ladam01} on the
shape of the phase function, which describes the angular dependence of
the scattered intensity for one scattering event.
\subsection{The multiple scattering of light by helium droplets} \label{atom:convolution}
The scattered intensity by a spherical droplet only depends on the
angle $\theta$ between the incident and the scattered directions, and
is described by the phase function $\Psi(u=$cos\,$\theta)$, the
probability distribution of $\theta$ after one scattering event,
multiplied by $4 \pi$.  The probability distribution $\Psi^{(p)}(u)$
after $p$ scattering events is obtained by successive convolutions of
$\Psi(u)$.  These convolutions amount to multiplications if $\Psi(u)$
and $\Psi^{(p)}(u)$ are expanded on Legendre polynomials~:
 \begin{equation}
    \Psi(u)=\sum_{l}^{\infty} (2 l +1) \nu_{l} P_{l}(u)
    \label{decompLegendre}
\end{equation}
\begin{equation}
    \Psi^{(p)}(u)=\sum_{l}^{\infty} (2 l +1) (\nu_{l})^{p} P_{l}(u)
    \label{decompLegendrep}
\end{equation}
with~: 
\begin{equation}
\nu _l=1/2 \int_{-1}^{1} {\Psi (u) P_l(u) du },
    \label{coeffLegendre}
\end{equation}

This allows to compute $\Psi^{(p)}(u)$ for any given initial 
distribution $\Psi(u)$\footnote{One should stress that this represents the 
angular distribution of the light, when summed over all possible positions 
after the $p$ scattering events, i.e. the exact equation~\ref{decompLegendrep} does not give 
any information on the spatial distribution of intensity, in contrast 
to the full radiative transfer equation\cite{vandehulst-book}.}.
A particular case is that of the phase functions $\Psi_{\alpha}(u,v)$ 
corresponding to coefficients $\nu_{l}$ given by~:

\begin{equation}
\nu _l=(v)^{l^{\alpha}}
    \label{coeffalpha}
\end{equation}
where $1\leq\alpha\leq 2$, and $v=<u>$, the average of cos\,$\theta$
over the distribution $\Psi$(cos\,$\theta$).  Note that
$1-v={l_{sca}}/{l^{*}}$, and that strong forward scattering
corresponds to $v$ close to 1.  These phase functions have a
functional form stable under the convolution process.  Special cases
are $\alpha$=1 and $\alpha$=2, corresponding respectively to the
Henyey-Greenstein (H-G) phase function \cite{HGorig}~:
\begin{equation}
\Psi_1(v,u)=\frac{(1-v^2)}{(1+v^2-2uv)^{3/2}} \label{eq:4}
\end{equation}
and to a nearly Gaussian distribution (the true Gaussian 
distribution  corresponds to $\nu_\ell=~v^{\ell(\ell+1)/2}$).
In both cases, the angular dependence after  $p$ scattering events is 
given by the same functional form,  with $v_{p}=v^p$. 

\begin{figure}
\center
\resizebox{1\columnwidth}{!}{%
  \includegraphics{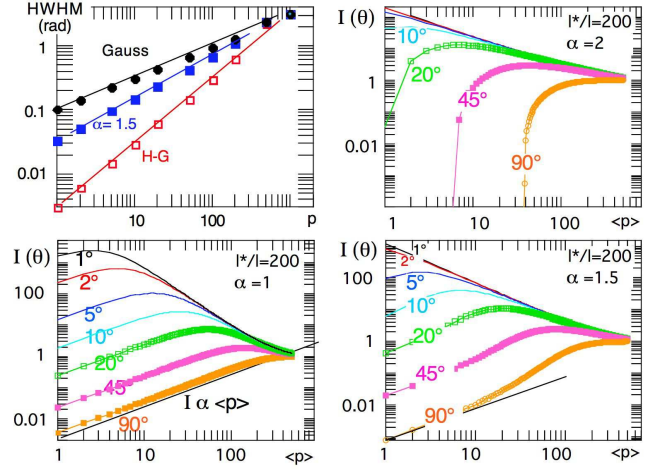}
 	}
\caption{(a) Half width at half maximum of the angular distribution 
of intensity, after $p$ scattering events, for three 
single-scattering phase functions 
of same r.m.s. width $<\theta^2>={2l_{sca}}/{l^{*}}$=0.01, differing by 
their values of $\alpha$ (see text). The straight lines correspond to 
a law $p^{1/\alpha}$. The widening of the angular 
distribution increases when $\alpha$ increases, i.e. when 
the original phase function has extended ``wings''; (b), (c), (d) :  
scattered intensity at different angles as a function of $<p>$ (see 
text) for the different values of $\alpha$. At large angles, the 
scattered intensity is initially linear in $<p>$ for  $\alpha$=1 and $\alpha$=1.5. This corresponds to a pseudo 
single-scattering regime. }
\label{theoryMS}       
\end{figure}

For small angular width and scattered angles, the H-G function is
closely similar to a 2D Lorentzian distribution around the forward
direction.  For a given half width at half maximum (HWHM)
$\theta_{HW}$, the large ``wings'' of such a distribution result in an
r.m.s. $\theta$ value much larger than $\theta_{HW}$,
$\theta_{rms}=<\theta^2>^{1/2}=({2l_{sca}}/{l^{*}})^{1/2}$, of order
$(\theta_{HW})^{1/2}$ This is in contrast with the Gaussian phase
function of same $\theta_{rms}$ for which the HWHM is of order
$\theta_{rms}$.  As first pointed out by Van de
Hulst\cite{vandehulst1996}, this implies that the H-G phase function
behaves very differently from the Gaussian phase function of same
$\theta_{rms}$ under multiple scattering.  In both cases, $v_{p}=v^p$
implies that $\theta_{rms}$ scales with $(p)^{1/2}$, but the
characteristic width for the H-G case increases linearly with $p$,
i.e. much faster than in the Gaussian case ($p^{1/2}$).  As shown by
figure~\ref{theoryMS}(a), for $\alpha=3/2$, intermediate between the
H-G and the Gaussian cases, $\theta_{HW}$ scales as $p^{2/3}$,
suggesting that the general behaviour is of the form
$\theta_{HW}\propto p^{1/\alpha}$, although we have not proven it
analytically.

Equation~\ref{decompLegendrep} also gives the behaviour at large
angles.  Figures~\ref{theoryMS}(b), \ref{theoryMS}(c),
\ref{theoryMS}(d) show how the intensity multiply scattered at various
angles $\theta$ varies with the average order of scattering $<p>$, for
a value of ${l^{*}/{l}}=1/(1-v)=200$, a ratio close to that for
helium, and respective values of $\alpha$ of 1, 1.5, 2.  Here, $<p>$
is the average number of scattering events, and the corresponding
multiply scattered phase function is obtained by summing
equation~\ref{decompLegendrep} over all $p$, with a weight given by
the Poisson distribution associated with $<p>$\cite{Ladam01}.  In all
cases, the intensity is transferred from small to large angles until
it is completely isotropized, but the way this isotropization takes
place strongly depends on the angular shape of the original phase
function, i.e. on $\alpha$.  In the Gaussian case, the intensity at
large angles only comes from the accumulation of small
single-scattering events : the larger the angle, the larger the number
of events needed to give some signal at this angle.  In contrast, for
the H-G case, the signal at large angles (compared to the HWHM of the
original phase function) initially grows linearly with $p$, whatever
the angle.  This implies that it comes from one large angle
single-scattering event separating two sequences of small-angle
scattering events, which nearly preserve the direction of propagation.
Because the large angle single-scattering event can occur at any
scattering step, the intensity is indeed proportional to $p$, the
total number of steps.  The range of validity of this regime increases
with the observation angle, and reaches about 200 scattering events at
90$^{\circ}$.  At this angle, a pseudo single-scattering behaviour can
be observed over a distance of order the transport mean path.  The
extension of the pseudo single-scattering regime depends on the shape
of the phase function.  Figure~\ref{theoryMS}(d) shows that this
regime still exists for $\alpha$=1.5, but in a smaller
range($<p>\,\leq$10, for angles $\theta \geq$20$^{\circ}$).

\begin{figure}
\center
\resizebox{0.95\columnwidth}{!}{%
  \includegraphics{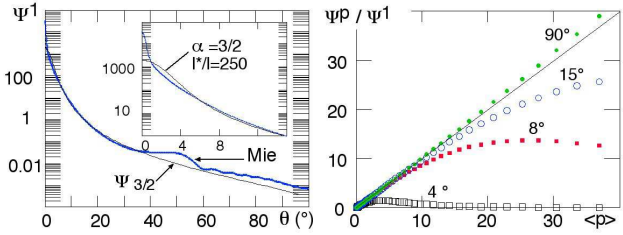}
} \caption{(a) Comparison of the Mie phase function for an exponential
distribution of diameters with a characteristic diameter of 20~$\mu$m
(at a wavelength 632 nm and for an index of refraction contrast of
1.025) to $\Psi_{3/2}$(cos $\theta$) for ${l^{*}/{l_{sca}}}$=250; (b)
Intensity multiply scattered at various angles as a function of the
average order of scattering $<p>$ for the Mie phase function.  Note
the existence of a linear regime up to 30 scattering events for a
scattering angle of $90^{\circ}$.}
\label{MieHe}       
\end{figure}

What about the real case of helium ?  Figure~\ref{MieHe}(a) shows the
Mie phase function for a realistic distribution of helium droplets.
The pronounced decay with angle is approximately accounted for by
classical refraction, below its maximal angle of deviation ($\approx
20^{\circ}$), and by reflections above, the bump around 45$^{\circ}$
being due to the rainbow effect.  The rise at small angles (inset) is
due to diffraction, or, more precisely, anomalous
scattering\cite{vandehulst}.  The Mie phase function is approximately
represented by $\Psi_{3/2}(u)$ for ${l^{*}/{l_{sca}}}$=250, so what we
expect a behaviour similar to that of figure~\ref{theoryMS}(d).  This
can be checked by numerical computation of the coefficients $\nu_{l}$
in this case and use of equation~\ref{decompLegendrep} to obtain the
intensity multiply scattered at various angles as a function of the
order of scattering $<p>$.  Figure~\ref{MieHe}(b) confirms our
expectation of a linear dependance in $<p>$, over a range which
increases with the observation angle.  In practice, we expect the
pseudo single-scattering effect to hold up to 10 scattering events for
an observation angle of $15^{\circ}$ (case of the sheet illumination)
and up to 30 scattering events for an observation angle of
$90^{\circ}$ (case of the vertical laser).

\begin{figure}
\center
\resizebox{\columnwidth}{!}{%
  \includegraphics{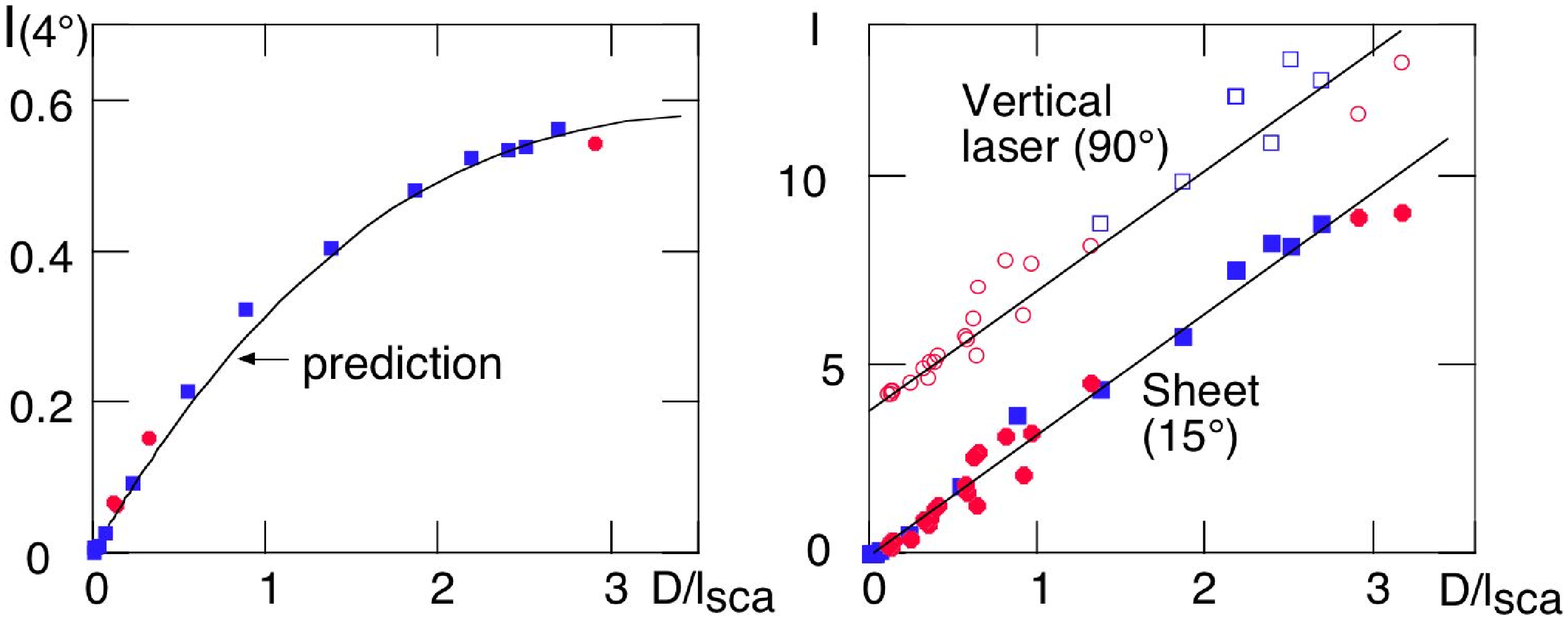}
} \caption{(a) Scattered intensity (arbitrary units) at 4$^{\circ}$ as
a function of the average number of scattering events $<p>$ deduced
from the attenuation of the beam.  The incident laser beam is
horizontal and probes the median plane of the tube.  Different symbols
correspond to different experimental conditions.  The continuous curve
is the predicted behaviour.(b) Scattered intensity at 15$^{\circ}$ and
90$^{\circ}$ on the pipe axis as a function of $<p>$.  Data for the
vertical laser beam are vertically shifted for clarity.  The linear
behaviour at $<p>$ larger than 1 is specifically due to the strong
forward scattering of helium.  }
\label{MSexp}       
\end{figure}
In our experiment, we have tested these calculations up to $<p>=3$ by
comparing the intensity scattered from the pipe horizontal median
plane to the average number of scattering events $<p>$ deduced from
the attenuation of the horizontal laser beam.  Figure~\ref{MSexp}(a)
shows that the intensity scattered at $\approx 4^{\circ}$ as measured
on the screen (figure~\ref{angulardep}) in not linear with $<p>$.
Its dependence is well accounted for by our calculation.  In contrast,
figure~\ref{MSexp}(b) shows that the intensity scattered at
$15^{\circ}$ and $90^{\circ}$ as measured on the images using sheet
and vertical laser illumination, respectively, linearly depends on
$<p>$, as expected.

\begin{figure}
\center
\resizebox{\columnwidth}{!}{%
 	    \includegraphics{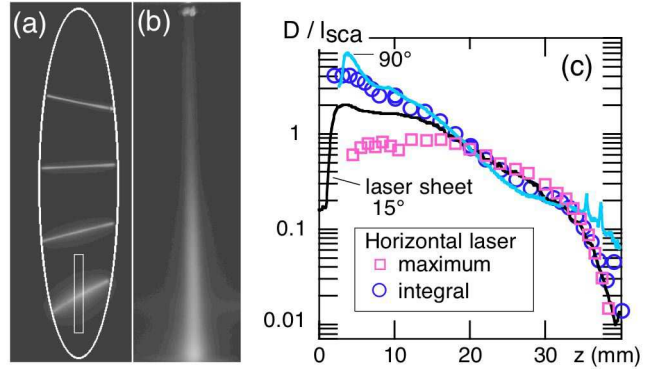}
	}
	\caption  {Comparison of the vertical profile of interfacial 
	density measured by different methods~:~laser sheet, and 
	vertically scanned horizontal laser beam observed at 
	$15^{\circ}$, vertical laser observed at 
	$90^{\circ}$. (a) is a superposition of four 
	images of the horizontal laser beam at four different 
	heights. In the bottom part of the tube, where 
	the droplets density is the largest, the beam is widened by multiple 
	scattering. The same effect affects the vertical beam (b). Correct measurements of the interfacial area 
	require to integrate the brightness over the beam width (box 
	in (a)) rather than to take its maximal value on the axis on the 
	beam. The results at  $15^{\circ}$ and $90^{\circ}$ are then 
	consistent (c).  
	}
	\label{graph:TravPDPA}
\end{figure}
Based on figure~\ref{MieHe}(b), we would expect the sheet illumination
to give the same interfacial area as the vertical laser for values of
$D/l_{sca}$ of order 10.  Figure~\ref{profilesV} shows that this is
not the case.  This disagreement is related to the fact that our
calculation predicts the total intensity scattered, but not where it
does come from.  Small scattering-angle events widen the incident
beam, which spreads the signal on the CCD over a larger region than
for simple scattering.  This effect is clearly visible on the
figure~\ref{graph:TravPDPA} in the bottom of the pipe, both with the
vertical laser beam, and with an horizontal one.  Thus, measuring the
local interfacial area from the brightness of a pixel on the axis of
the horizontal laser beam leads to underestimate it.  As shown by
figure~\ref{graph:TravPDPA}(c), the larger the interfacial area (i.e.
the closer to the bottom of the pipe), the larger the underestimation.
The error made with the sheet illumination is less than with the
horizontal laser beam, because the spreading of light outside a given
pixel is partly compensated by the spreading from lower or larger
heights.  However, unlike with the sheet, a valid measurement of the
interfacial area remains possible with the horizontal laser beam by
integrating the optical signal over the laser whole effective width,
as shown in figure~\ref{graph:TravPDPA}(c) (such an integration was
also performed to obtain the points of figure~\ref{profilesV}(b)).  In
this way, the interfacial area deduced from the scattering at
$15^{\circ}$ is consistent with that deduced at $90^{\circ}$.

To summarise this section, we have shown that meaningful
measurements of the interfacial intensity are possible with helium, 
even in a regime of quite strong scattering, provided that one observes at a 
large angle enough and one takes into account the widening of the 
beam due to multiple scattering. This advantage of helium was 
crucial to our optical measurements of developed atomisation.

\subsection{Atomisation threshold and stratification}\label{atom:results}
The combination of optics with the 400~W refrigerator capabilities
enabled us to study the dependence of the atomisation process on three
parameters~:~the level of liquid in the pipe, the vapour velocity, and
the temperature, the latter setting the density of the vapour, and the
state of the liquid (superfluid below 2.17~K, normal above).  The
results\cite{ThesePerraud} will be described
elsewhere\cite{Perraud2008}.  In particular, we have characterised two
properties which determine the potentialities of the atomised flow for
refrigeration, the velocity threshold for atomisation, and, above this
threshold, the mist stratification.  In the following, we will
describe these properties and compare them to those for the water-air
system, so as to pinpoint the specificities of helium.
\begin{figure}
\center
\resizebox{0.9\columnwidth}{!}{%
 \includegraphics{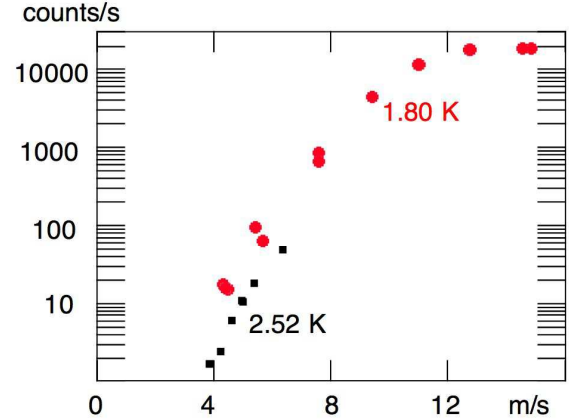}}	
\caption {The vapour velocity dependence of the PDPA count rate gives
an upper bound of 4~m/s for the atomisation threshold.  The section
probed by the PDPA (about 150 $\times$ 500 $\mu$m) is located on the
pipe axis.  The velocity is the average droplet velocity, as measured
by the PDPA, which is about 20\% larger than the average of the vapour
velocity over the pipe section, computed from the vapour total
flow rate.  The total helium flow is 14.5~g/s at 1.8~K, and 16~g/s at
2.52~K, corresponding to a bulk liquid level at 4.5~m/s of
approximately 35\% the pipe perimeter at 1.8~K, and 24\% at 2.52~K.
Despite the larger vapour density at 2.52~K, the count rate is lower,
due to the lower level, combined with the stronger stratification.}
\label{threshold}
\end{figure}

Let us first discuss the atomisation threshold, and its sensitivity to
the superfluidity of the liquid.  Figure~\ref{threshold} shows how the
the PDPA counting rate increases with the vapour velocity at constant
temperature and total flow-rate, both below and above the superfluid
transition (1.8~K and 2.52~K).  For both temperatures, droplets are
detected on the pipe axis above the minimal velocity used, about
4~m/s.  The exponential decrease of the counting rate at small
velocities does not allow to extrapolate a velocity threshold from
these measurements.  The atomisation threshold was thus determined
from pictures of the whole pipe section using sheet illumination.
Droplets were detected above 3.4$\pm$0.4~m/s at 1.8~K, and
2.5$\pm$0.3~m/s at 2.52~K. These velocities are two to three times the
Kelvin-Helmoltz threshold for wave formation\cite{Guyon91}, $U_{KH}=
({4 \sigma g \rho_{L} }/{\rho_{V}^{2}})^{1/4}$, where $g$ is the
gravity acceleration, and the other quantities are defined in
table~\ref{tableHe}.  $U_{KH}$ is 1.7~m/s at 1.8~K et 0.8~m/s at
2.52~K. They can also be compared to the Ishii-Grolmes
criterion\cite{Ishii75} for atomisation, which takes into account the
effect of viscosity in a phenomenological way.  It predicts a
threshold of about 2.2~m/s at both temperatures, the smaller vapour
density at 1.8~K being compensated for by the smaller viscosity.
While it remarkably accounts for the threshold in the normal state,
the Ishii-Grolmes criterion underestimates the threshold in the
superfluid state, and, worse, predicts an atomisation threshold below
the threshold for wave formation, showing that it overestimates the
influence of a vanishing viscosity.  In any case, our experiment shows
that the influence of superfluidity on the atomisation threshold is at
most limited.  The threshold is lower than for the water-air system
(where it typically lies in the range 10-15~m/s, depending on the pipe
diameter and the liquid flow rate \cite{Ishii75,Williams96}), but this
can be mainly ascribed to the small density and surface tension of
liquid helium, the product of which is about 1/2000 of that for water.
As this product enters $U_{KH}$ with a power 1/4, we also understand
why the difference between helium and water remains nevertheless
limited.

Above the atomisation threshold, a stratified mist is observed.
Measurements at constant velocity show that the temperature (in the
superfluid region) and liquid level do affect the degree of
atomisation, but not the profile of stratification.  This allows to
study the effect of velocity on stratification by comparing
experiments at different temperatures and flow rates.

\begin{figure}
\center
\resizebox{\columnwidth}{!}{%
 \includegraphics{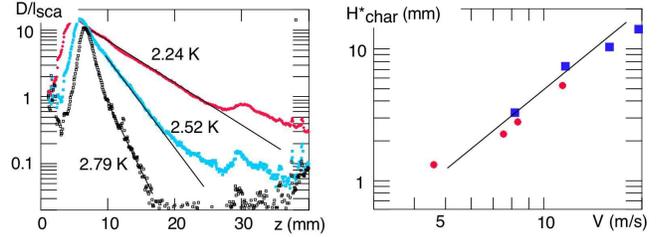}}	
\caption {(a) Stratification for normal fluid, as measured using the
vertical laser.  The applied power is 360~W, and the total helium flow
rate is 23~g/s.  The liquid level covers from 22\% to 24\% of the
perimeter.  The inverse of the slope of the straight lines is the
stratification height $H_{char}$.  (b) Velocity dependence of 
$H_{char}^{*}$=$\eta_{v} (1.8~K) H_{char}
/\eta_{v} (T)$, which cancels the theoretical dependence
of $H_{char}$ on the vapour viscosity, for points at 1.8~K (squares) and
in the normal phase (circles).  The vapour velocity $V_{v}$ is measured by the
PDPA on the pipe axis.  The straight line corresponds to a $V_{v}^2$ 
dependence.}
\label{stratifnormale}
\end{figure}

Figure~\ref{stratifnormale}(a) shows the stratification profile for
three temperatures in the normal region, as deduced from the intensity
scattered at $90^{\circ}$ from the vertical laser, which we have shown
in \S\ref{atom:convolution} to be the least sensitive to multiple
scattering effects.  The applied power being constant, the vapour
velocity decreases with increasing vapour density, hence temperature.
As in the superfluid phase, this implies a stronger stratification.
For both figures~\ref{profilesV} and \ref{stratifnormale}, the profile
is approximately exponential, so that we can quantify the
stratification by the characteristic length $H_{char}$ such that
$\Sigma(z) \propto$ exp$ (-z/H_{char})$.  Figure~\ref{stratifnormale}(b)
shows $H_{char}$, measured above 10~mm, for the profiles of
figures~\ref{profilesV} and \ref{stratifnormale}, as a function of
the vapour velocity $V_{v}$ measured on the pipe axis with the PDPA.
$H_{char}$ is found to follow an approximate $V_{v}^{2}$ dependence,
similar to what we have extracted from droplets concentration profiles
measured for the water-air system\cite{Pan02}.

For droplets of diameter $d$ small enough to follow the gas
turbulence, $H_{char} \approx D_{turb}/V_{fall}$, the ratio of the
turbulent diffusivity $D_{turb}$ to the droplets fall velocity
$V_{fall} \propto d^2/\eta_{v}$.  For isotropic turbulence, $D_{turb}$
is expected to scale with the vapour velocity, so that $H_{char} \propto
\eta_{v} V_{v}/d^2$. $H_{char}/\eta_{V}$ vary
approximately as $V_{v}^{2}$, suggesting that the droplets diameter
decreases with the vapour velocity as $d \propto V_{v}^{-1/2}$,
independent on the superfluid nature of the liquid.  Also, since the
stratification, at a fixed vapour velocity, does not depend on
temperature, the droplet diameter should only weakly depend on the vapour
density.

The decrease of diameter with increasing velocity is physically
expected, although measurements in the water-air system generally find
a stronger $V_{v}^{-1}$ dependence\cite{Simmons01}.  However, this
conclusion is at variance from the results of direct measurements of
the diameter distribution on the pipe axis using the
PDPA\cite{ThesePerraud,Perraud2008}.  They show a nearly exponential
size distribution, with an average diameter increasing with the vapour
velocity (approximately from 20 to 40 $\mu$m).  Measurements at
temperatures between 1.8 and 2.0~K show that the diameter dependence
on vapour density follows a similar trend.  Both points are
inconsistent with the above analysis of the stratification.  The
reason for this discrepancy remains to be understood.

Despite this problem, the range of average diameters given by the 
PDPA ($\approx 10-30 \mu$m) is consistent with the angular dependence 
of the intensity scattered at small angle (figure~\ref{angulardep}), and, 
independently, on the value of $\Sigma$, combined with the 
concentration of droplets, as inferred from the PDPA count rate and 
velocity. These diameters are smaller than reported for the water-air 
interface. For example, for the latter system, Simmons and Hanratty 
\cite{Simmons01}
report average diameters on the pipe axis decreasing from 60 to 35 
$\mu$m, for air velocities increasing from 30 to 50 m/s (i.e. two to three times 
the atomisation threshold). Here again, the 
difference probably stems from the very low surface tension of helium.

\section{The case of small scatterers : condensation and evaporation
of helium inside mesoporous glasses } \label{condensation}
\subsection{Background} \label{background}
The phenomenon of condensation and evaporation of fluids in mesoporous
media, i.e. with pores sizes falling in the nm-$\mu$m range, is
studied as an example of a phase transition in presence of disorder
and confinement\cite{Gelb99}, and used to characterise pore
distributions.  Both the condensation and evaporation processes
involve the formation of domains of liquid or vapour on a microscopic
scale.  The associated fluctuations of optical index scatter light so
that it is attractive to study the phenomenon in transparent porous
media, such as silica glasses (Vycor, silica aerogels, \ldots).
However, in some instances, the length scale of the fluctuations can
reach several hundreds of nanometers, impeding single-scattering
measurements for usual fluids.  This problem can be solved by using
helium as a fluid.  Figure~\ref{figurelpm} shows that, for helium
droplets of diameter 200~nm, the mean free path remains larger than
one millimetre even for an extrapolated volume fraction of 50\%,
allowing conventional light scattering measurements on a millimetre
thick sample.  This would not be possible for, e.g., water, where the
mean free path for a similar distribution of droplets would be 200
times smaller.  We have taken advantage of this specificity of helium
to address several fundamental problems in the field using optical
means.

The processes of condensation and evaporation are usually
characterised by sorption isotherms, which measure the adsorbed amount
of fluid as a function of increasing (for condensation) or decreasing
(for evaporation) pressure.  In the case where the dense phase of the
fluid wets the substrate, one observes the reversible adsorption of a
thin film at low vapour pressure, followed by capillary condensation,
a rather abrupt filling at a pressure of order, but smaller, than the
saturated vapour pressure.  The later process is hysteretic, emptying
occurring at a lower pressure than filling.  A central question is the
origin of this hysteresis.  Different explanations invoke the
metastability of the gas-like phase\cite{BallEvansEL87}, changes in
the shape of menisci between condensation and
evaporation\cite{SaamColePRB75}, collective effects for networks of
connected pores such as Vycor\cite{Mason88}, and, more recently, the
energetic and geometric disorder of the porous media\cite{Kierlik01}.
A characteristic feature of the two last explanations is that
condensation and evaporation differ in a fundamental way, which is the
influence of nucleation.  Unlike condensation, which can proceed from
the adsorbed film, evaporation requires the nucleation of vapour in
the dense phase, which involves an activation energy.  If this process
cannot occur, a pore cannot empty as long as its neighbours remain
filled, so that desorption should take place through a collective
percolation process starting from the surfaces of the sample.

These different explanations for hysteresis predict different shapes
for the hysteresis loop.  For example, the percolation mechanism
should manifest through a sharp kink in the desorption isotherm at the
percolation threshold\cite{Mason88}, followed by a nearly vertical
portion, as observed in Vycor and other materials.  Also, for weak
enough disorder, the disorder-based mechanism predicts a change of
shape, from smooth to steep, of the adsorption isotherm as the
temperature is decreased below a critical value\cite{Kierlik01}.
Still, measuring the shape of the loop only is ambiguous, and optical
measurements allow to test more precisely the different mechanismes
proposed.  This is beautifully illustrated by the work of Page \it{et
al}\rm, who studied the condensation and evaporation of hexane in
Vycor using light scattering\footnote{Note that, in this experiment,
results were limited to pressures very close to the percolation
threshold.  For smaller pressures, multiple scattering was so large
that no light was transmitted, in agreement with the discussion at the
beginning of~\S\ref{background}.}, and found long-range correlations
in the vapour distribution, characteristic of a percolation process,
close to the kink of the desorption isotherm\cite{PagePRE95}.
However, up to now, there were no systematic study of the correlation
between the shape of the isotherms, and the distribution of fluid
inside a porous medium, both at the microscopic and macroscopic
levels.  This is the aim of our experiments.  In the following, we
will illustrate the potential of our technique by several examples.
\subsection{Experimental} \label{experimental}
\begin{figure}
\center
\resizebox{\columnwidth}{!}{%
 	    \includegraphics{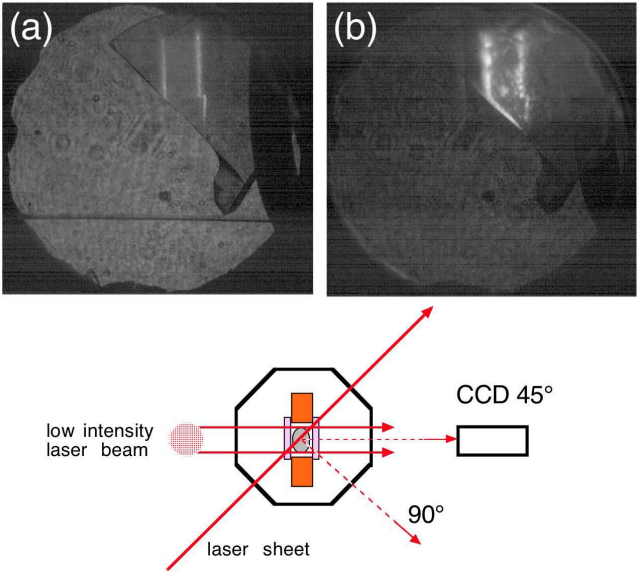}
} \caption {Illustration of the scattering geometry.  The sample is a
disk, or (here) a fragment of a disk, illuminated at 45$^\circ$ from
its faces by a thin laser sheet.  For these pictures, the background
is illuminated with a weak extended laser beam through the viewport
normally used for observation at 135$^\circ$.  (a) corresponds to the
filled aerogel (the bulk interface is visible below the aerogel), and
(b) to the beginning of the desorption process.  In the filled state,
scattering is only due to the silica backbone, immersed in liquid
helium at an uniform density.  The interception lines of the laser
sheet with the back and front surfaces are brighter , due to forward
scattering by these surfaces.  During desorption, white regions
appear, due to the coexistence of liquid and vapour on a microscopic
scale.}
	\label{vueglobaleB100}
\end{figure}
We study different porous glasses, silica aerogels and Vycor, using an
optical cryostat with optical ports 45$^\circ$ apart.  The sample
cell, 20~mm in diameter and 4~mm thick, has two sapphire windows,
allowing observation under different scattering angles. Sorption
isotherms are measured between 4.2~K and the critical bulk temperature
($\approx$5.2 K), as described in \cite{Lambert04} and \cite{Cross07}.  Simultaneously,
the distribution of helium inside the sample is determined by light
scattering measurements.  To that aim, the sample is illuminated by a
thin He-Ne laser sheet under a 45$^{\circ}$ incidence with respect to
its faces, and imaged at different angles (45$^\circ$, 90$^\circ$, and
135$^\circ$) using CCD cameras, as illustrated on
figure~\ref{vueglobaleB100}.

\subsection{Evidence for a disorder-driven transition in silica aerogels} \label{disorderdriven}
Recent numerical studies\cite{Kierlik01,Sarkisov02}, based on a
mean-field density functional theory, suggest that hysteresis could
result from the disorder of the porous media.  They predict that a
disorder-driven transition could occur as a function of disorder or
temperature, similar to that occurring in the Random Field Ising Model
at zero temperature \cite{Sethna}.  This out-of-equilibrium phase
transition implies a change of shape, from smooth to steep, of the
condensation branch of the hysteresis loop, when the porosity is
increased at a constant temperature or, alternatively, when the
temperature is decreased below some critical value, at a constant
porosity\cite{DetcheverryE03}.

\begin{figure}
\center
\resizebox{\columnwidth}{!}{%
 	    \includegraphics{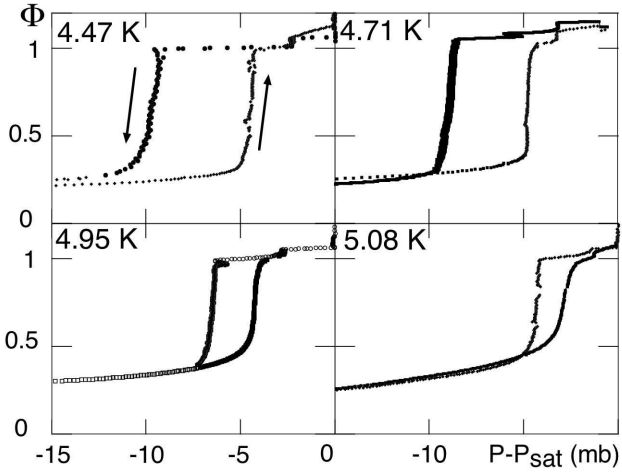}
}	
	\caption  {Isotherms for a base-catalyzed aerogel of porosity 
	95\% (B100). The adsorbed fraction in aerogel, $\Phi$, is 
	plotted as a function of the pressure, referred to the bulk saturated 
	vapour pressure.  The scanning time is typically one day for 
	each branch of the loop. The adsorption branch evolves from steep to 
	smooth as the temperature is increased, whereas, for 
	desorption, it is nearly vertical at all temperatures.}
	\label{isosB100}
\end{figure}

\begin{figure}
\center
\resizebox{\columnwidth}{!}{%
 	    \includegraphics{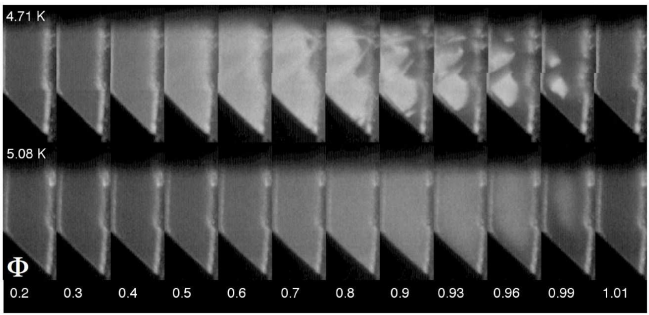}
	} \caption {Images of aerogel B100, observed at 45$^\circ$
	from the incident laser sheet, for increasing adsorbed
	fractions $\Phi$.  The condensation morphology is correlated
	to the steepness of the adsorption isotherm 
	(figure~\ref{isosB100}).}
	\label{imagescondensation}
\end{figure}

We have recently provided the first evidence for this transition in a
base-catalyzed silica aerogel\cite{Bonnet08}.  In this sample
(denominated B100), the silica forms a complex arrangement of
interconnected strands, resulting in a large porosity (95\%) and an
associated weak disorder.  For this porosity, the recent theoretical
studies, performed on aerogels numerically synthesised by Diffusion
Limited Cluster Aggregation\cite{DetcheverryE03}, predict the
occurrence of a disorder-driven transition.  As shown by
figure~\ref{isosB100}, the adsorption isotherms become steeper at low
temperatures, which is indeed in agreement with the scenario of a
disorder-driven transition, and cannot be explained by the usual
description of the capillary condensation.  The corresponding optical
observations are shown in figure~\ref{imagescondensation}.  At low
adsorbed fraction $\Phi$ ($\Phi < 50-60$\%), the scattered intensity
increases uniformly, corresponding to the development of liquid
domains on a microscopic scale (up to typically 200~nm, from the
absolute value of the scattered intensity and its angular dependence).
Above $\Phi = 50-60$\%, the behavior depends on temperature.  For the
low temperature points (corresponding to the steep isotherms), the
injected helium completely fills some regions of the aerogel, making
them dark.  The size of these dark regions increases up to the point
where the whole aerogel is filled. In contrast, at temperatures where the isotherm is smooth,
the condensation proceeds in an homogeneous way.  As discussed in
detail in ref.\cite{Bonnet08}, these observations are consistent with
the occurrence of a disorder-driven phase transition.

\subsection{Absence of long-range correlations during desorption} 
\label{desorption}
\begin{figure}
\center
\resizebox{1\columnwidth}{!}{%
 	    \includegraphics{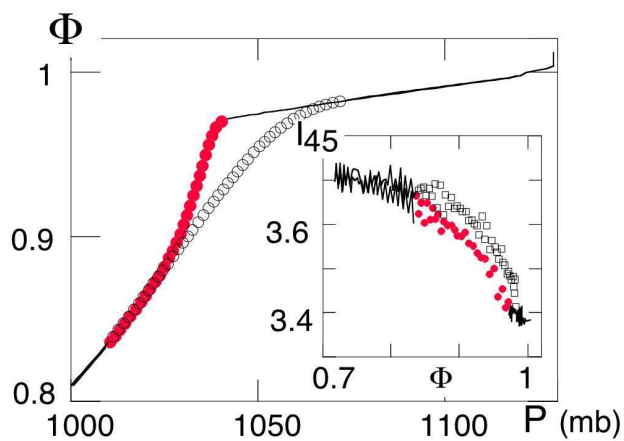}
	} \caption { Hysteresis loop for Vycor at
	4.34~K. Note that hysteresis occurs over a small range of 
	$\Phi$. The inset shows the scattered intensity at 
	45$^{\circ}$ as a function of $\Phi$. Along the hysteresis 
	loop, scattering is slightly larger for adsorption
	(squares) than it is for desorption (circles).}
	\label{Vycor}
\end{figure}

\begin{figure}
\center
\resizebox{1\columnwidth}{!}{%
 	    \includegraphics{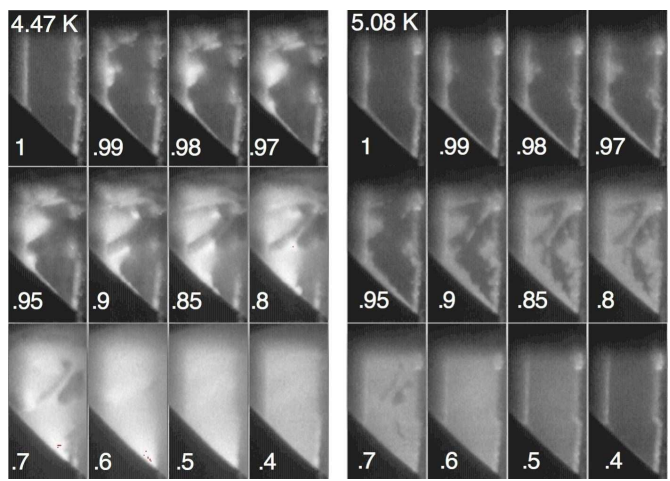}
	} \caption {Images of desorption for aerogel B100, at 4.47~K
	and 5.08~K, for decreasing liquid volume fractions $\Phi$.
	Desorption starts from the aerogel surfaces.  }
	\label{imagesVidangeB100}
\end{figure}

\begin{figure}
\center
\resizebox{1\columnwidth}{!}{%
 	    \includegraphics{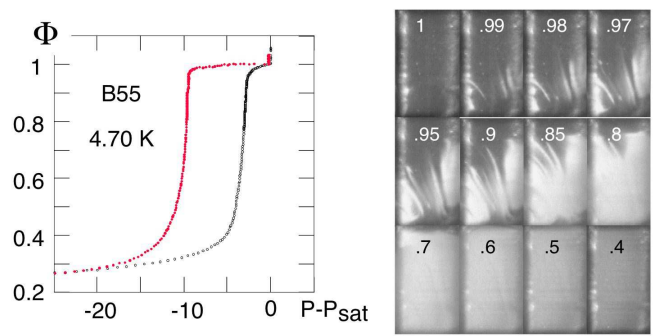}
	} \caption {Hysteresis loop for aerogel B55 at 4.70~K and
	corresponding images during the desorption process.  As in
	B100, desorption starts abruptly and close to the surfaces.  }
	\label{imagesVidangeB55}
\end{figure}
We have also studied the behavior during desorption\cite{Bonnet08b}.
In order to compare to the experiment of Page \it{et al}\rm, we have
studied a sample of Vycor.  At the lowest temperature presently
achievable with our set-up (4.3~K), the desorption isotherm does not
have an initial vertical portion (figure~\ref{Vycor}), in contrast
with the situation at lower temperatures\cite{Brewer62}.  The
scattered intensity decreases monotonously from the empty state to the
filled state.  This means that, unlike in aerogels, the initial
adsorption of the film decreases, instead of increases, the scattered
intensity.  This is related to the difference in porosities.  For the
aerogels, the silica concentration is small (several \%).  Adding more
silica or a film of helium on the surface of the filaments increases
the scattered field.  In contrast, for Vycor, the silica concentration
is large (70\%), so that one might see the scattering as originating
from the cavities.  Reducing the cavity size decreases the scattered
signal, and so does the addition of an helium film.  More
surprizingly, the scattered signal never exceeds its value in the
empty state, even along the hysteresis loop where it is slightly
larger for adsorption than for desorption (inset of
figure~\ref{Vycor}).  This implies that no microdomains are formed on
the scale of the hundred of nanometers.  Moreover, there is no
evidence for the long-range correlations during desorption which would
sign a percolation process.  On the other hand, the kink in the
desorption isotherm is not a sharp one, so that our Vycor experiment
does not rule out the usual hypothesis of a correlation between the
desorption mechanism and the shape of the desorption isotherm.

What about aerogels? In this case, the desorption iso-therms present a
vertical portion for all temperatures (figure~\ref{isosB100}).  The
same holds for the two other aerogels described in
ref.\cite{Bonnet08}, which have different microstructures and
porosities.  Based on the above hypothesis, one would
conclude that a percolation process is at play.
Figure~\ref{imagesVidangeB100} shows pictures of the desorption
process for the previous sample.  When the adsorbed fraction starts to
decrease (corresponding to the kink in the isotherms), bright regions
appear, demonstrating the apparition of vapour ``bubbles'' at the
microscopic scale.  In the initial stage of emptying, these regions
are located close to the surfaces of the aerogel, (i.e. near the
bright lines marking the back and front surfaces, and on the bottom
part of the aerogel).  The same conclusions apply to the other
aerogels.  For example, figure~\ref{imagesVidangeB55} shows the
desorption process for a base-catalyzed sample of porosity 97.5\%
(B55).  Desorption initially occurs along some ``fingers'', some of
which distinctly originate from the aerogel surfaces.  This points to
a specific role of surface during the desorption process, seemingly consistent
with the percolation scenario.

\begin{figure}
\center
\resizebox{1\columnwidth}{!}{%
 	    \includegraphics{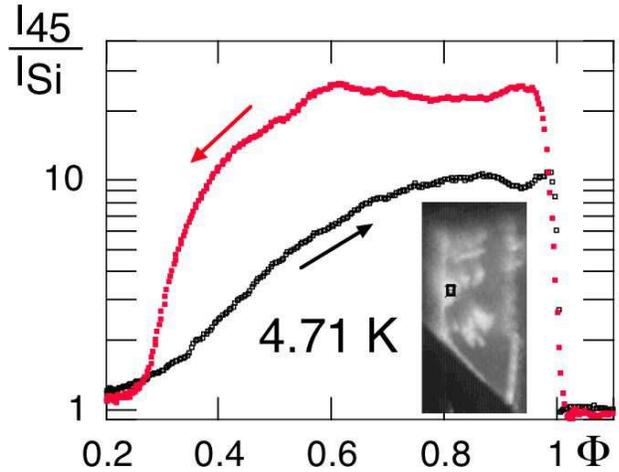}
	} \caption { Intensity scattered at 45$^{\circ}$ for B100.
	The intensity is the average grey level in the rectangle of
	the picture, normalized by the silica background in the same
	rectangle.  The difference between adsorption and desorption
	is moderate, implying that the correlations during desorption
	do not extend on much larger scales than during adsorption.  }
	\label{I45B100}
\end{figure}
\begin{figure}
\center
\resizebox{1\columnwidth}{!}{%
 	    \includegraphics{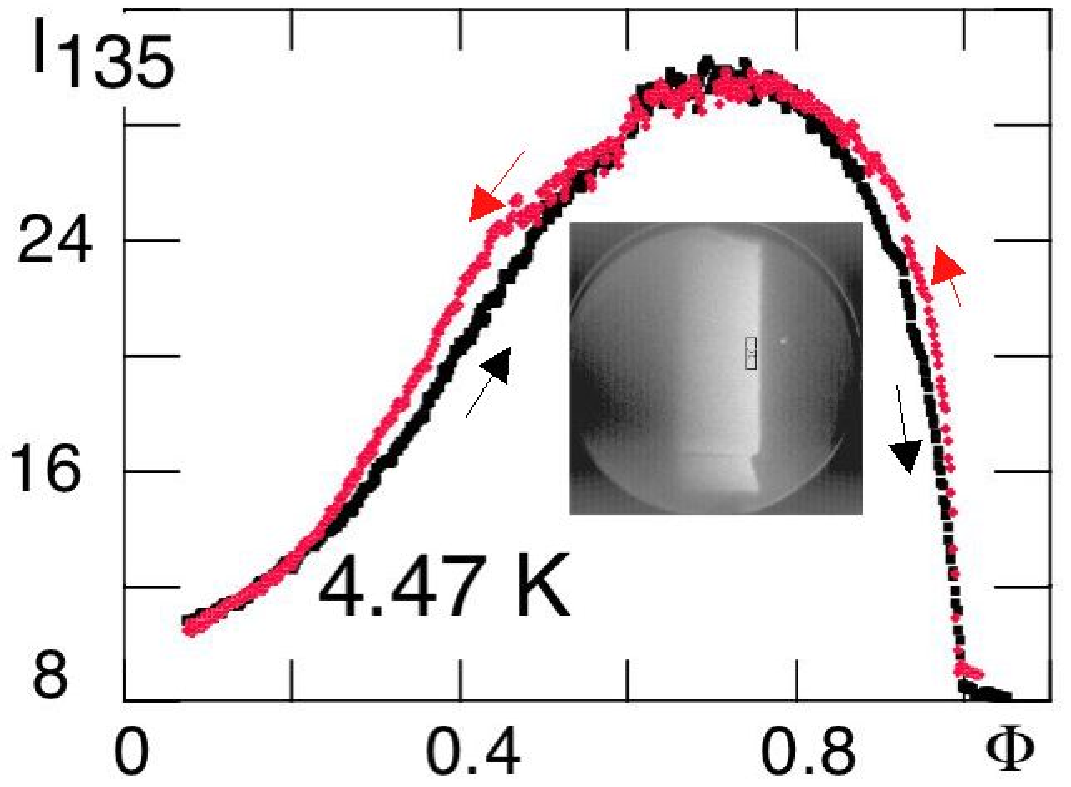}
	} \caption {Intensity scattered at 135$^{\circ}$ for N102.
	Due to its larger correlation length, the empty aerogel N102
	scatters light much more strongly than B100 or B55.  The
	scattering mean free path is comparable to the sample
	thickness, so that the light backscattered at 135$^{\circ}$ is
	significantly attenuated, depending on the probed depth.  To
	avoid the complication of this effect, the scattered intensity
	is measured close to the entrance surface of the beam.  When
	plotted as a function of $\Phi$, it is nearly identical
	between adsorption and desorption.  This is valid as well for
	other points in the sample, showing that, on the scale of the
	hundred of nanometers, the distribution of liquid is
	unexpectedly the same for adsorption and desorption.}
	\label{I135N102}
\end{figure}
In this scenario, one expects a large increase of scattering with
respect to the case of condensation, as observed for hexane in
Vycor\cite{PagePRE95}.  However, for all three samples studied, we
found no evidence for such an increase.  This is illustrated by
figures~\ref{I45B100} and \ref{I135N102}.  Figure~\ref{I45B100} shows,
for the base-catalyzed sample B100, the intensity scattered at
45$^{\circ}$ from a point close to the surface, where the bright
regions first appear during desorption.  At a given average adsorbed
fraction, the signal during evaporation is somewhat larger than during
condensation, but by a factor less than 3, which can be
explained\cite{Bonnet08b} by a modest increase in size (typically
40\%) of the correlated domains.  The case of the neutrally-catalysed
sample (N102) is even more spectacular.  Despite a sharp kink in the
desorption isotherm, the optical signal shown in figure~\ref{I135N102}
is nearly reversible as a function of $\Phi$.  The same results hold
for all temperatures, and for B55, which means that, for all three
aerogels, desorption does not involve much longer length scales than
adsorption\footnote{This contrasts with the results of neutron
scattering studies for aqueous gels\cite{Li94}.  However, the length
scales probed in that case did not exceed a few tens of nm, an order
of magnitude smaller than those we probe.}.  To our knowledge, these
results are the first evidence that the kink in the desorption
isotherm does not necessarily follow from a percolation process.

\begin{figure*}
\center
\resizebox{2\columnwidth}{!}{%
 	    \includegraphics{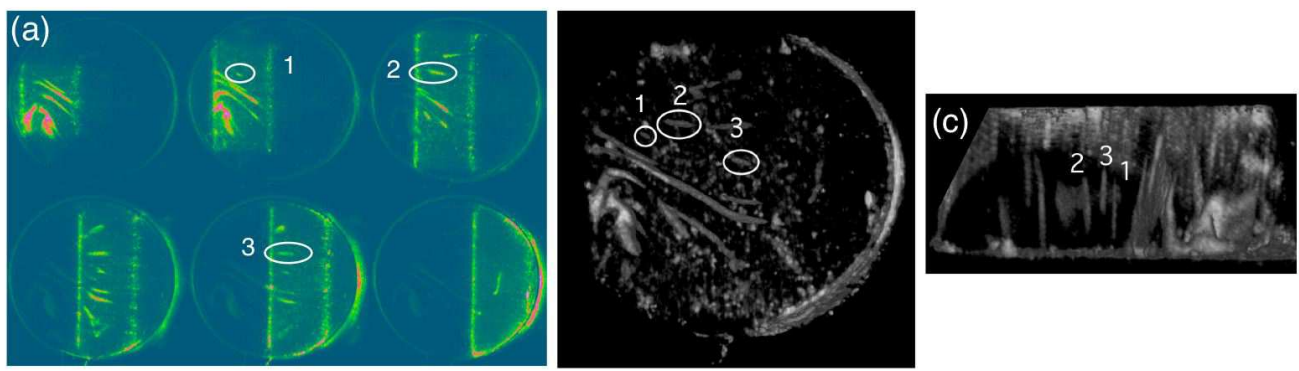}
}	
	\caption  {Evidence for bulk nucleation during desorption in aerogel 
	B55: (a) Several slices at constant liquid volume fraction 
	and temperature; (b) Reconstructed volume seen parallel to the 
	surface of the disk. The circled regions do not contact the 
	outer perimeter of the disk; (c) View orthogonal to (b), 
	showing that the circled regions do not contact the disk 
	faces either (top and bottom on (c)). (b) and (c) are 
	reconstructed using the ImageJ plug-in Volume Viewer.
	 }
	\label{tomoH97}
\end{figure*}
\subsection{Testing the occurrence of nucleation} \label{nucleation}
The absence of long-range correlations for B100 contrasts with the
predictions of the mean-field density functional approach.
Numerical calculations of the structure factor for a base-catalysed
aerogel of porosity 87\% predict that the scattering during desorption
should be much larger than during adsorption\cite{DetcheverryE06}.  A
possible explanation would be that, in the real aerogel, thermal
activation, which is neglected in the numerical studies, could allow
to overcome the nucleation barrier.  If this were true, one should
observe the apparition of vapour ``bubbles'', hence of bright regions,
in the bulk of the sample.  For testing this issue, a single slice of
the sample is not enough.  For example, the bright ``fingers'' which
appear in the middle of the laser sheet in
figure~\ref{imagesVidangeB55} could be connected to the surfaces
outside of the sheet plane.  Detecting the occurrence of nucleation
thus requires to reconstruct the whole 3D sample.  This is possible by
scanning the laser sheet parallel to itself and capturing the
resulting images.  In this way, one obtains a stack of parallel slices
of the sample, which can be reconstructed using the
freeware ImageJ \cite{ImageJ}.  Figure~\ref{tomoH97} shows B55, soon
after the beginning of desorption.  Several small bright regions can
be identified as non-connected to the surfaces.  If confirmed, this
result will be an evidence for nucleation.  Light scattering could
then be used to study how the nucleation mechanism depends on the
temperature and the microstructure of samples.

\section{Conclusions}
In this paper, we have shown how the weak optical contrast of helium
droplets or bubbles allows to perform quantitative light scattering
measurements on diphasic systems at large volumic fractions.  We have
exploited this property to study optically the atomisation of liquid
helium and the gas-liquid phase transition in porous media.  The
original results obtained so far illustrate the richness of blending
low temperature and soft matter physics, and will be a motivation to
pursue this approach further.

\section{Acknowledgments}
We thank B. Cross, L. Guyon, T. Lambert, and E. Di 
Muoio for having contributed to our earlier results, and R. van 
Weelderen and P. Lebrun for useful discussions. We are grateful to 
CERN and ANR-06-BLAN-0098 for having respectively supported the Cryoloop and 
aerogels experiments.

\end{document}